\documentclass{Interspeech2024}

\interspeechcameraready 
\usepackage{cite}
\usepackage{hyperref}
\usepackage{url}
\usepackage{multirow}
\usepackage{tabularx}
\newcommand\numberthis{\addtocounter{equation}{1}\tag{\theequation}}

\title{Relational Proxy Loss for Audio--Text based Keyword Spotting}

\name{Youngmoon}{Jung}
\name{Seungjin}{Lee}
\name{Joon-Young}{Yang}
\name{Jaeyoung}{Roh}
\name{Chang Woo}{Han}
\name{\\Hoon-Young}{Cho}

\address{
  Samsung Research, Seoul, South Korea}
\email{\{youngm.jung, sjsr.lee, jy610.yang, jyo.roh, cw1105.han, h.y.cho\}@samsung.com}


\begin{document}

\maketitle

\begin{abstract}
    
    In recent years, there has been an increasing focus on user convenience, leading to increased interest in text-based keyword enrollment systems for keyword spotting (KWS). Since the system utilizes text input during the enrollment phase and audio input during actual usage, we call this task audio--text based KWS.  
    To enable this task, both acoustic and text encoders are typically trained using deep metric learning loss functions, such as triplet- and proxy-based losses. 
    This study aims to improve existing methods by leveraging the structural relations within acoustic embeddings and within text embeddings.
    Unlike previous studies that only compare acoustic and text embeddings on a point-to-point basis, our approach focuses on the relational structures within the embedding space by introducing the concept of Relational Proxy Loss (RPL).
    By incorporating RPL, we demonstrated improved performance on the Wall Street Journal (WSJ) corpus.
    
\end{abstract}

\section{Introduction}

Keyword spotting (KWS) aims at identifying pre-specified keywords in audio streams. Recently, KWS has attracted considerable research attention with the popularity of voice assistants that are triggered by keywords such as ``Alexa,'' ``Hi Bixby,'' or ``Okay Google.'' 
KWS techniques can be broadly classified into two categories: fixed KWS \cite{Chen14-ICASSP, Sainath-INTERSPEECH, TANG17-ICASSP} and flexible (or user-defined) KWS \cite{Chen-ICASSP, Huang-ICASSP, Kurmi-INTERSPEECH, Lim-arxiv, He-ICLR, Jung-INTERSPEECH, Jung-INTERSPEECH-AdaMS, Shin-INTERSPEECH, Nishu-INTERSPEECH, Lee-INTERSPEECH}.
Unlike fixed KWS where users are required to utter fixed keywords exclusively, flexible KWS allows users to enroll and speak arbitrary keywords.
Specifically, in flexible KWS, user-defined or customized keywords can be enrolled either in the form of audio \cite{Chen-ICASSP, Huang-ICASSP, Kurmi-INTERSPEECH, Lim-arxiv} or text \cite{He-ICLR, Jung-INTERSPEECH, Jung-INTERSPEECH-AdaMS, Shin-INTERSPEECH, Nishu-INTERSPEECH, Lee-INTERSPEECH}.
Because the text-based keyword enrollment does not require multiple utterances of a target keyword, but can be easily achieved through text typing, the demand for text-enrolled flexible KWS is increasing.

Text-enrolled flexible KWS systems \cite{He-ICLR, Jung-INTERSPEECH, Jung-INTERSPEECH-AdaMS, Shin-INTERSPEECH, Nishu-INTERSPEECH, Lee-INTERSPEECH} typically utilize a text encoder during the enrollment phase and an acoustic encoder during the test phase,
where both encoders are optimized employing deep metric learning (DML) \cite{Wang-CVPR} objectives such as contrastive loss \cite{Nishu-INTERSPEECH}, triplet-based loss \cite{He-ICLR}, and proxy-based loss \cite{Jung-INTERSPEECH, Jung-INTERSPEECH-AdaMS}.
As per \cite{Wang-CVPR}, DML aims at learning an embedding space where the embedding vectors of similar samples are encouraged to be closer, while dissimilar ones are pushed further apart.
Particularly in text-enrolled flexible KWS, acoustic and text embeddings representing the same keyword must become closer, while those representing different keywords must not.
Considering that the acceptance or rejection of a keyword is determined via measuring the similarity between audio and text inputs, we refer to this task as audio--text based KWS.

Previous research on audio--text based KWS has focused on two main aspects:
1) audio--text alignment and 2) DML loss function. 
In \cite{Shin-INTERSPEECH} and \cite{Lee-INTERSPEECH}, cross- and self-attention modules were employed, respectively, to extract audio--text alignment information (termed ``pattern'').
Subsequently, a binary classifier was trained with cross-entropy loss to determine if the aligned audio and text inputs represent the same keyword \cite{Shin-INTERSPEECH, Lee-INTERSPEECH}.
Nishu \textit{et al}. \cite{Nishu-INTERSPEECH} proposed Dynamic Sequence Partitioning algorithm for audio--text alignment, where the aligned audio and text embeddings were trained using the contrastive loss function.
Regarding the DML framework,
He \textit{et al}. \cite{He-ICLR} proposed the multi-view learning approach that jointly learns acoustic and text embeddings using triplet-based loss.
In \cite{Jung-INTERSPEECH}, the multi-view learning method was extended to a proxy-based DML by employing the text embeddings as proxies, which was further improved in \cite{Jung-INTERSPEECH-AdaMS} by introducing learnable scale and margin parameters used in the DML framework.

\begin{figure}[t]
  \centering
  \includegraphics[width=\columnwidth]{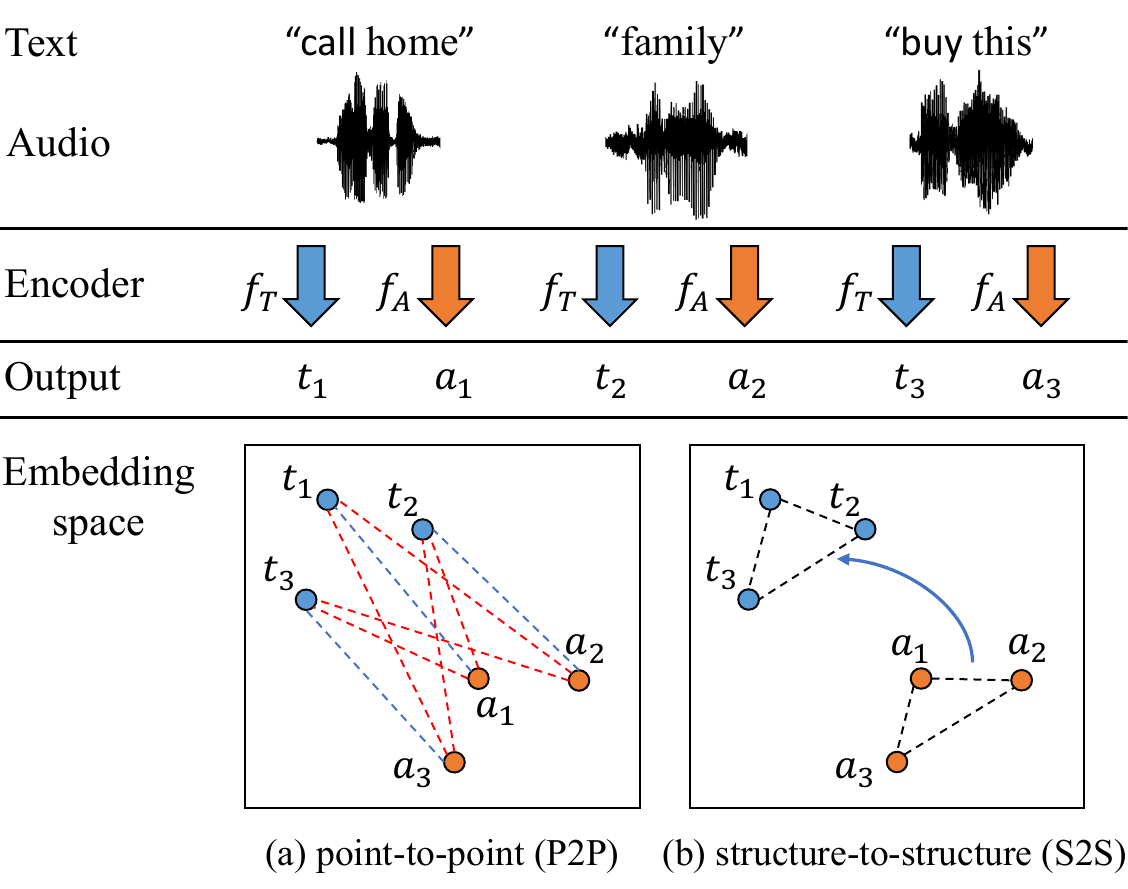}
  \vspace{-0.5cm}
  \caption{Illustration of Relational Proxy Loss. $f_T$ and $f_A$ denote text and acoustic encoders that generate text embedding $t_i$ and acoustic embedding $a_i$, respectively, where $i$ is the sample index. In (a), the red and blue dashed lines indicate negative and positive relations between points. In (b), black dashed line denotes structural information.}
  \label{fig:fig1_overview}
  \vspace{-0.5cm}
\end{figure}

\begin{figure*}[t]
  \centering
  \includegraphics[width=\textwidth]{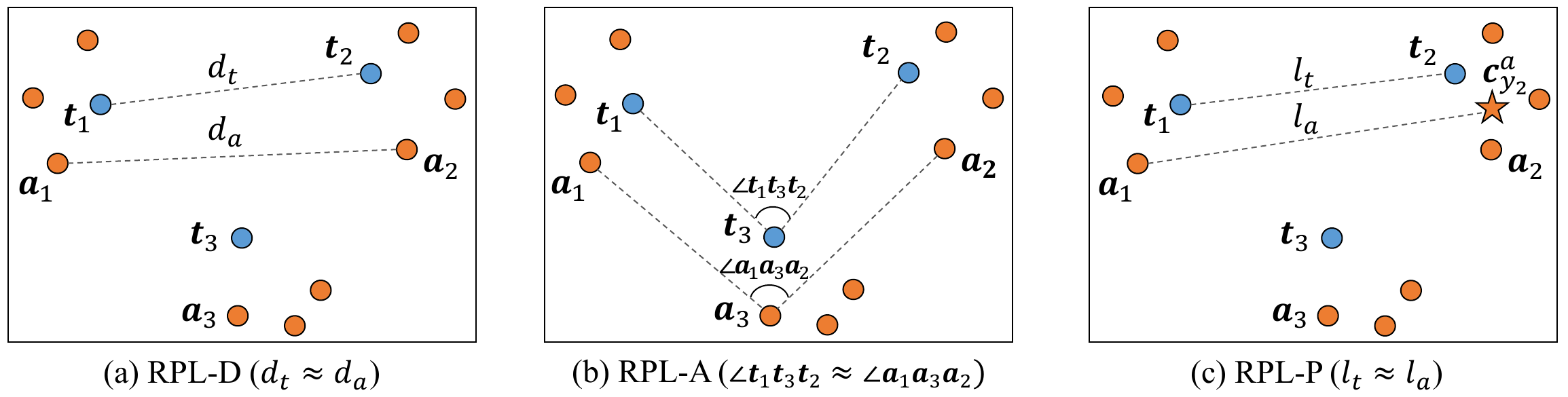}
  \vspace{-0.5cm}
  \caption{Illustration of three types of Relational Proxy Loss (RPL) in the embedding space: (a) distance-wise RPL (RPL-D), (b) angle-wise RPL (RPL-A), and (c) Prototypical RPL (RPL-P). The notations in Fig. \ref{fig:fig1_overview} are maintained here. $d$ and $l$ represent the distances between embeddings used in RPL-D and RPL-P, respectively. $\bm{c}_{y_2}^a$ refers to the prototype of the class to which the $\bm{a}_2$ belongs.}
  \label{fig:fig2_RPL}
  \vspace{-0.4cm}
\end{figure*}

In this paper, we focus on the DML loss function and propose Relational Proxy Loss (RPL). We leverage structural relations of acoustic embedding (AE) and text embedding (TE) motivated by Park \textit{et al}.'s work \cite{Park-CVPR}. The authors highlighted the importance of the structure in the embedding space, which constitutes a vital component of the knowledge learned by a trained model.
We incorporate this concept in the context of DML by employing the RPL, which utilizes the relational information within AEs and TEs.
Similar to \cite{Jung-INTERSPEECH}, we treat TEs to proxies. 
The primary information of proxies representing the corresponding word classes is well-presented by the structural relations, so we can assume that the AEs belonging to the corresponding classes are expected to follow the same relations.
Fig. \ref{fig:fig1_overview} illustrates the distinction between (a) traditional DML-based approaches and (b) our proposed RPL. 
While conventional DML losses compare an AE and a TE by computing their similarity (point-to-point), the proposed RPL compares structural relations within the embeddings (structure-to-structure).
Through the implementation of RPL, the structural relations of AEs are adjusted and brought closer to those of TEs, thereby enhancing the comparability between AE and TE.
Specifically, we compute distance- and angle-wise relations for both AE and TE independently. We then combine the point-to-point and structure-to-structure approaches, yielding better overall system performance than either approach alone.
We evaluate performance on Wall Street Journal (WSJ) corpus. 
Experimental results show that the proposed approach outperforms previous approaches under noisy and reverberant conditions.

\section{Proposed method}
\label{section:Proposed_method}

We assume that a mini-batch comprises a set of $N$ data tuples, which is denoted as $\mathcal{X} = \{(\bm{a}_i, \bm{t}_i, y_i)|i=1,2,\cdots,N\}$. Here, $\bm{a}_i$ is the acoustic embedding (AE) of the $i$-th speech segment, $\bm{t}_i$ is the text embedding (TE) of the corresponding text, and $y_i \in \{1,\cdots,K\}$ is the corresponding word label.

\subsection{Proxy-based losses}
\label{section:AsyP}
We begin by discussing proxy-based methods, which serve as our baselines of the point-to-point (P2P) approach in Fig. \ref{fig:fig1_overview}(a).
The asymmetric-proxy (AsyP) loss \cite{Jung-INTERSPEECH} is defined as the sum of the anchor--positive and anchor--negative terms, as follows:
\begin{align*}
    \mathcal{L}_\textrm{AsyP}=\frac{1}{N} \sum_{i=1}^N \bigg (\frac{1}{\alpha} \underset{j \in \mathcal{P}_i}{\textrm{ELSE}} \, \alpha(\lambda - S(\bm{t}_i,\bm{a}_j)) \\ + \underset{k \in \mathcal{N}_i}{\textrm{MSP}} \, \beta(S(\bm{a}_i,\bm{t}_k) - \lambda) \bigg), \numberthis \label{eq1}
\end{align*}
where $\mathcal{P}_i = \{j|y_j=y_i\}$, $\mathcal{N}_i = \{k|y_k \neq y_i\}$, and $S(\cdot,\cdot)$ are the set of positive indices, the set of negative indices, and the cosine similarity, respectively. $\textrm{ELSE}$ and $\textrm{MSP}$ denote the Extended-LogSumExp \cite{Wang-CVPR} and Mean-Softplus \cite{Yi-ICPR} functions, respectively.
The first term brings anchor $\bm{t}$ and positive examples $\bm{a}$ closer together, while the second term separates the anchor $\bm{a}$ and negative examples $\bm{t}$. 
Although the TE is utilized as a proxy, it contributes to the second term as a negative sample rather than acting as an anchor.
Here, $\alpha$, $\beta$, and $\lambda$ are hyper-parameters that dictate the boundaries in the embedding space or the degree of penalty for violations. Although they are vital elements during training, their values need to be manually optimized.
In \cite{Jung-INTERSPEECH-AdaMS}, the Adaptive Margin and Scale (AdaMS) was suggested, which defines these hyper-parameters as learnable parameters specific to each keyword class.

\subsection{Relational Proxy Loss}

Motivated by \cite{Jung-INTERSPEECH}, we treat TEs as proxies representing distinct word classes because each TE is generated for a specific word class.
The primary goal of the Relational Proxy Loss (RPL) is to transfer structural information from TEs to AEs by exploiting the relationships among TEs. 
By applying the RPL, we train the acoustic encoder to adopt the same relational structure as that of the text encoder through a relational potential function $\phi$ that quantifies the relational energy of the given $n$-tuple. 
Here, we denote $\mathcal{X}_{a}^n$ as the set comprising all possible $n$-tuples of $\bm{a}$ derived from $\mathcal{X}$, e.g., $\mathcal{X}_a^2 = \{(\bm{a}_i,\bm{a}_j)|i \neq j \}$ and $\mathcal{X}_a^3 = \{(\bm{a}_i,\bm{a}_j,\bm{a}_k)|i \neq j \neq k \}$. 
Likewise, $\mathcal{X}_t^n$ is defined as the set encompassing all possible $n$-tuples of TEs obtained from $\mathcal{X}$.

First, we establish two types of potential functions: a distance-wise function $\phi_{D}$ using pairwise examples $\mathcal{X}_a^2$ and $\mathcal{X}_t^2$, and an angle-wise function $\phi_{A}$ using ternary examples $\mathcal{X}_a^3$ and $\mathcal{X}_t^3$.
Based on $\phi_{D}$ and $\phi_{A}$, we define distance-wise RPL ($\textrm{RPL-D}$) and angle-wise RPL ($\textrm{RPL-A}$), respectively. Both are visualized in Fig. \ref{fig:fig2_RPL}(a) and (b), respectively.
To begin with, the $\textrm{RPL-D}$ can be formulated as follows:
\begin{equation}
    \mathcal{L}_\textrm{RPL-D} = \sum_{\substack{(\bm{t}_i,\bm{t}_j) \in \mathcal{X}_t^2 \\  (\bm{a}_i,\bm{a}_j) \in \mathcal{X}_a^2}} h_{\delta}(\phi_D(\bm{t}_i,\bm{t}_j),\phi_D(\bm{a}_i,\bm{a}_j)), \label{eq2} 
\end{equation}
where $h_{\delta}$ represents a loss function that aims to minimize the difference of the relational energies measured via $\phi$ between the text and acoustic encoders, expressed as Huber loss \cite{Huber-AMS} with a scaling factor of 1.
Note that $h_{\delta}$ is also used in defining other types of RPL losses.
For a given pair of samples, $\phi_{D}$ computes the Euclidean distance between two instances in the embedding space:
\begin{equation}
    \phi_D(\bm{a}_i,\bm{a}_j) = \frac{1}{\mu_a} ||\bm{a}_i-\bm{a}_j||_2, 
\end{equation}
where $\mu_{a}$ denotes the average distance between pairs of $\bm{a}$, serving as a normalization factor.
We define $\mu_{a}$ as follows:
\begin{equation}
    \mu_{\bm{a}} = \frac{1}{|\mathcal{X}_{a}^2|} \sum_{(\bm{a}_i,\bm{a}_j) \in \mathcal{X}_{a}^2} ||\bm{a}_i - \bm{a}_j||_2.
\end{equation}
$\phi_D(\bm{t}_i,\bm{t}_j)$ is defined in the same way as $\phi_D(\bm{a}_i,\bm{a}_j)$.
Through this method, the acoustic encoder is guided towards focusing on the distance structures of AEs.

Given a triplet $\mathcal{X}_a^3$ or $\mathcal{X}_t^3$, an angle-wise potential function $\phi_{A}$ calculates the angle created by the three instances in the embedding space. To simplify computation, we opt for calculating the cosine value of the angle, which can be represented using an inner product, rather than directly obtaining the angle value: 
\begin{align*}
    \phi_A(\bm{a}_i,\bm{a}_j, \bm{a}_k) = \textrm{cos}\angle \bm{a}_i\bm{a}_j\bm{a}_k = \frac{\bm{u}_{ij} \cdot \bm{u}_{kj}}{\Vert\bm{u}_{ij}\Vert \Vert\bm{u}_{kj}\Vert},
    \numberthis \label{eq5}
\end{align*}
where $\bm{u}_{ij}=\bm{a}_i-\bm{a}_j$ and $\bm{u}_{kj}=\bm{a}_k-\bm{a}_j$.  
We aim to quantify the angle between $\bm{u}_{ij}$ and $\bm{u}_{kj}$, denoted by $\angle \bm{a}_i\bm{a}_j\bm{a}_k$. 
Considering this, we define the $\textrm{RPL-A}$ as follows:
\begin{equation}
    \mathcal{L}_\textrm{RPL-A} = \sum_{\substack{(\bm{t}_i,\bm{t}_j,\bm{t}_k) \in \mathcal{X}_t^3 \\  (\bm{a}_i,\bm{a}_j,\bm{a}_k) \in \mathcal{X}_a^3}} h_{\delta}(\phi_A(\bm{t}_i,\bm{t}_j,\bm{t}_k),\phi_A(\bm{a}_i,\bm{a}_j,\bm{a}_k)).
\end{equation}
The $\textrm{RPL-A}$ transfers relational knowledge about angles from the text encoder to the acoustic encoder, encouraging the acoustic encoder to focus on the angle structures of AEs.

Finally, we introduce an extra kind of distance-based RPL that makes use of prototypes (i.e., class centroids). Adopting the concept from \cite{Snell-NIPS}, we define the prototype of the AE for the class $k$ as follows:
\begin{equation}\label{eqn:eq7}
    \bm{c}_k^a = \frac{1}{|\mathcal{X}_k|} \sum_{\bm{a}_i \in \mathcal{X}_k } \bm{a}_{i},
\end{equation}
where $\mathcal{X}_k$ denotes the subset of examples from $\mathcal{X}$ belonging to class $k$.
Since each keyword class is represented by a distincitive TE, the TE itself corresponds to the class centroid.
Nevertheless, to simplify the notation during the loss term definition, we also apply Eq. (\ref{eqn:eq7}) to the TE.
The final loss, called Prototypical RPL (RPL-P), is defined as follows:
\begin{equation}\label{eqn:eq8}
    \mathcal{L}_\textrm{RPL-P} = \sum_{\bm{x}_i \in \mathcal{X}} \sum_{k=1}^K h_{\delta}(\phi_D(\bm{t}_{i},\bm{c}_k^t),\phi_D(\bm{a}_{i},\bm{c}_k^a)),
\end{equation}
where $\bm{x}_i = (\bm{a}_{i}, \bm{t}_{i}, y_i) \in \mathcal{X}$.
By doing so, the distance between each embedding and the set of $K$ class centroids is computed.
As shown in Fig. \ref{fig:fig2_RPL}(c), the RPL-P transfers distance information between proxies (i.e., TEs) to the distance between the AE and its corresponding class centroid.

In the above, it is assumed that the classes of TEs are distinct from one another, as this is commonly observed within a mini-batch setting.
When the classes coincide, the RPLs behave similarly to other types of losses. 
For example, RPL-D in Eq. (\ref{eq2}) resembles the cross-entropy loss used for AE classification, since RPL-D pulls $\bm{a}_i$ and $\bm{a}_j$ closer together due to $\phi_D(\bm{t}_i,\bm{t}_j)$ being equal to zero.
Also, RPL-P in Eq. (\ref{eqn:eq8}) is akin to the Prototypical loss \cite{Snell-NIPS} if the class label $y_i$ matches $k$ and the value of $\phi_D(\bm{t}_{i},\bm{c}_k^t)$ becomes zero. 
As such, the AE moves towards its corresponding centroid because of the term $\phi_D(\bm{a}_{i},\bm{c}_k^a)$.

\section{Experiments}
\subsection{Experimental setup}

We used the King-ASR-066 \cite{Speechocean-DB} dataset which contains recordings from 2.6k native English speakers for training.
Each sample was recorded at a 16 kHz sampling rate in home/office environments.
All utterances were divided into word-level segments by forced alignment using the Montreal Forced Aligner \cite{McAuliffe-INTERSPEECH}, and we made use of 4.6k hours of speech data. The training set comprised phrases containing one to five words, with a total of 211,676 classes.
For data augmentation, we convolved the clean speech signals with synthetic room impulse responses (RIRs) from the OpenSLR dataset \cite{Ko-ICASSP} and added various kinds of noise, such as babble, car, and music, at randomly selected signal-to-noise ratios (SNRs) between -3 and 25 dB.

To evaluate the performance of all systems under consideration, we used the Wall Street Journal (WSJ) \cite{Paul-WSN} corpus, acquired using the WSJ recipe available within the Kaldi toolkit \cite{Povey-ASRU}.
For development and testing, we employed the following subsets of the dataset: \{dev93, dev93-5k\} and \{eval92, eval92-5k, eval93, eval93-5k\}.
These development and testing sets contained a total of 16,839 and 18,274 word-level segments, containing 3,289 and 3,239 unique words, respectively.
To simulate real-world scenarios, we generated noisy and reverberant speech by convolving synthetic RIRs from the OpenSLR dataset and adding the MUSAN noise dataset \cite{Snyder-arxiv} with SNRs varying from 5 to 25 dB. 
As a result, we obtained development and test sets with the same amount of data as their respective original sets.
During evaluation, we randomly selected 1.3M positive and 13M negative pairs from the test set.
We measured the model performance using two metrics: the Equal-Error-Rate (EER) \cite{Shin-INTERSPEECH, Lee-INTERSPEECH, Nishu-INTERSPEECH} and the Average Precision (AP) \cite{He-ICLR, Jung-INTERSPEECH, Hu-INTERSPEECH, Jung-ASRU}, both of which are commonly used in KWS.
As input features, we used 40-dimensional log Mel-filterbank coefficients with a frame length of 25 ms and a frame shift of 10 ms, which were mean-normalized throughout each utterance.

All experiments were conducted using the TensorFlow library.
During the training, the AdamW optimizer was employed with an initial learning rate of $10^{-4}$, which was halved every 20th epoch.
A weight decay parameter of $10^{-5}$ was applied throughout the training, which involved a total of 100 epochs. The entire training process required roughly one day to complete on two A100 GPUs.
Each mini-batch consisted of 500 utterances derived from 250 different keywords, with 2 utterances per keyword.
For the AsyP loss in Eq. (\ref{eq1}), we fixed $\alpha = 2$, $\beta = 50$, and $\lambda = 0.1$. These values served as the initial values for their corresponding parameters in the AdaMS.

For the acoustic encoder, we adopted the ECAPA-TDNN \cite{Desplanques-arxiv} with 256 filters in the convolutional layers (2.2M parameters).
After statistics pooling, a 512-dimensional AE was obtained. 
In the text encoder, we incorporated a pre-trained byte pair encoding (BPE) \cite{Sennrich-ACL} tokenizer and a two-layer bi-LSTM with 512 hidden units (10M parameters). 
We used the implementation of a character-level tokenizer available at HuggingFace \cite{MoiHuggingFace-Tokenizers} and trained the model with a vocabulary of 100 tokens on an internal dataset consisting of 50M English texts. 
The tokenizer split the input text into a sequence of subword units, and then the resulting sequence was transformed into a 1024-dimensional vector sequence by a trainable look-up table. The 1024-dimensional vector sequence was fed into the following bi-LSTM.
At each time step, the outputs of the bi-LSTM were concatenated and then passed through a global average pooling, followed by a fully connected layer, producing a 512-dimensional TE.
The design choices for these architectures were determined based on the computational resources available for deployment on our target devices.



\begin{table}[]
\renewcommand\thetable{1}
\caption{Ablation study for Relational Proxy Loss (RPL).}
\vspace{-0.1cm}
\label{tab:tab1}
\footnotesize   
\centering
\begin{tabular}{c|ccc|c}
\hline
\multirow{2}{*}{P2P} & \multicolumn{3}{c|}{S2S} & \multirow{2}{*}{AP (\%)}  \\ \cline{2-4}
                        & \multicolumn{1}{c|}{RPL-D} & \multicolumn{1}{c|}{RPL-A} & RPL-P                         &                         \\ \hline
\checkmark                       & \multicolumn{1}{c|}{-}     & \multicolumn{1}{c|}{-}     & -     & 71.9                          \\ 
\checkmark                       & \multicolumn{1}{c|}{\checkmark}     & \multicolumn{1}{c|}{-}     & -     & 76.7                                 \\ 
\checkmark                      & \multicolumn{1}{c|}{-}     & \multicolumn{1}{c|}{\checkmark}    & -     & 77.3                                 \\  \checkmark
                        & \multicolumn{1}{c|}{\checkmark}      & \multicolumn{1}{c|}{\checkmark}      & -     & 77.0                               \\ 
\checkmark                       & \multicolumn{1}{c|}{\checkmark}   & \multicolumn{1}{c|}{-}     &   \checkmark    & 78.3                                 \\ 
\checkmark                       & \multicolumn{1}{c|}{-}     & \multicolumn{1}{c|}{\checkmark}      &   \checkmark    & 77.2                                     \\ 
-                       & \multicolumn{1}{c|}{\checkmark}   & \multicolumn{1}{c|}{\checkmark}      &  \checkmark     & 66.8                                   \\ \hline\hline
\checkmark                       & \multicolumn{1}{c|}{\checkmark}   & \multicolumn{1}{c|}{\checkmark}      &   \checkmark    & \textbf{79.4}                                    \\ \hline
\end{tabular}
\vspace{-0.2cm}
\end{table}

\subsection{Ablation study for Relational Proxy Loss (RPL)}
This paper uses `P2P' and `S2S' to represent `point-to-point' and `structure-to-structure' methods, respectively, as shown in Fig. \ref{fig:fig1_overview}.
To ensure a fair comparison, all systems share the same experimental setup, including datasets and model architectures, but differ only in their training criteria.

First, we conduct an ablation study to verify the effectiveness of each RPL loss, as presented in Table \ref{tab:tab1}. In this case, we designate P2P as the AdaMS method, combining the Adaptive Margin and Scale (AdaMS) into the AsyP loss, as explained in Section \ref{section:AsyP}. This serves as the baseline in the table, showing 71.9\% of Average Precision (AP). 
We observe that three RPL losses (i.e., RPL-D, RPL-A, and RPL-P) improve the performance of the P2P method.
When we incorporate RPL-D and RPL-A into P2P, we achieve 76.7\% and 77.3\% of AP.
Upon adding RPL-A and RPL-P to `P2P + RPL-D', we can obtain additional improvements, yielding 77.0\% and 78.3\% of AP, respectively.
On the other hand, when we incorporate RPL-D and RPL-P into `P2P + RPL-A', we are not able to achieve any improvements for both scenarios.
Nevertheless, upon comparing the system with an AP of 79.4\% and one with an AP of 78.3\%, RPL-A enhances the performance of `P2P + RPL-D + RPL-P'. 
In summary, Integrating all RPL losses leads to a noticeable improvement in the performance of the `P2P' method, resulting in an AP of 79.4\% with a relative improvement (RI) of 10.43\% compared to the baseline.
Additionally, we investigate the performance when only three RPL losses are employed without P2P. 
As a result, the performance declines, yielding 66.8\% of AP. 
Thus, based on these findings, we can conclude that employing both P2P and S2S demonstrates the better performance than either approach alone, and using all four losses shows optimum performance among all systems presented in the table.

\subsection{Effects of auxiliary losses}
\label{section:aux_losses}
Table \ref{tab:tab2} shows the effects of auxiliary losses (AL) on the performance of our proposed system. Here, `RPL' refers to our system using P2P with three RPL losses in Table \ref{tab:tab1}.
The initial loss is referred to as the prototype--centroid matching loss, denoted as $\mathcal{L}_{\textrm{pc}}$. It is computed based on the centroids obtained through Eq. (\ref{eqn:eq7}), representing the average Euclidean distance between each prototype (i.e., TE) and its corresponding centroid of AE within a mini-batch.
As illustrated in Fig. \ref{fig:fig2_RPL}(c), $\mathcal{L}_{\textrm{pc}}$ makes the distance between $\bm{t}_2$ and $\bm{c}_{y_{2}}^a$ become zero. 
Since both are representative embeddings that represent each word class, we aim to bring them closer using $\mathcal{L}_{\textrm{pc}}$.

The losses discussed so far are all related to the relationship between AE and TE, regardless of whether it is P2P or S2S. 
However, we find that incorporating losses from the speech-enrolled KWS approach is beneficial for the text-enrolled KWS as well. 
Specifically, we introduce losses from \cite{Lim-arxiv}, consisting of two auxiliary losses, namely $\mathcal{L}_{\textrm{mono}}$ and $\mathcal{L}_{\textrm{triplet}}$, exclusively associated with the acoustic encoder.
In \cite{Lim-arxiv}, the acoustic encoder is trained through multi-task learning, employing triplet loss ($\mathcal{L}_{\textrm{triplet}}$) for word-level classification and cross-entropy loss ($\mathcal{L}_{\textrm{mono}}$) for frame-level monophone classification. 
To compute $\mathcal{L}_{\textrm{mono}}$, we insert a softmax layer taking the frame-level features as input, which are obtained before performing global average pooling. 
For $\mathcal{L}_{\textrm{triplet}}$, we use the 512-dimensional AE as the input.
By leveraging these auxiliary losses, we improve the performance of the proposed RPL, achieving an AP of 82.5\% and demonstrating their effectiveness in amplifying the capability to discern subtle differences between classes.


\begin{table}[t]
\renewcommand\thetable{2}
\caption{Effect of auxiliary losses (AL) for RPL.}
\vspace{-0.2cm}
\label{tab:tab2}
\footnotesize   
\centering
\begin{tabular}{c|c|c|c|c}
\hline
\multicolumn{1}{l|}{RPL} & \multicolumn{1}{l|}{$\mathcal{L}_{\textrm{pc}}$} & \multicolumn{1}{l|}{$\mathcal{L}_{\textrm{mono}}$} & \multicolumn{1}{l|}{$\mathcal{L}_{\textrm{triplet}}$} &  \multicolumn{1}{l}{AP (\%)} \\ \hline
\checkmark                           & -                          & -                          &                -          &      79.4                \\
\checkmark                           & \checkmark                          & -                          &              -            &            80.9             \\ 
\checkmark                           & -                        & \checkmark                          &       \checkmark                   &        81.0                 \\ \hline\hline
\checkmark                           & \checkmark                          & \checkmark                           &           \checkmark               &        \textbf{82.5}                               \\ \hline
\end{tabular}
\vspace{-0.1cm}
\end{table}

\begin{table}[t!]
\renewcommand\thetable{3}
\caption{Comparison of the performance between the proposed methods and other state-of-the-art systems.}
\vspace{-0.25cm}
\label{tab:tab3}
\footnotesize   
\centering
\begin{tabular}{c|c|c|c|c|c}
\hline
\multirow{1}{*}{Method} & \multirow{1}{*}{\begin{tabular}[c]{@{}c@{}}Enroll type\end{tabular}} & \multirow{1}{*}{P2P} & \multirow{1}{*}{S2S} & \multirow{1}{*}{EER (\%)} & \multirow{1}{*}{AP (\%)} \\ \hline
PATN \cite{Lim-arxiv}                 & Speech                                                                 & \checkmark                       & -                         & \multicolumn{1}{c|}{7.58}        & 74.5        \\
AsyP \cite{Jung-INTERSPEECH}                 & Text                                                                   & \checkmark                       & -                          & \multicolumn{1}{c|}{7.04}        & 71.3        \\ 
AdaMS \cite{Jung-INTERSPEECH-AdaMS}                 & Text                                                                   & \checkmark                       & -                           & \multicolumn{1}{c|}{6.99}       & 71.9       \\ \hline\hline

RPL                   & Text                                                                   & \checkmark                       & \checkmark                             & \multicolumn{1}{c|}{6.30}       & 79.4       \\ 
\textbf{RPL+AL}                 & Text                                                                   & \checkmark                       & \checkmark                          & \multicolumn{1}{c|}{\textbf{5.84}}       & \textbf{82.5}       \\ \hline
\end{tabular}
\vspace{-0.3cm}
\end{table}

\subsection{Comparison with the state-of-the art methods}
In Table \ref{tab:tab3}, we compare the performances of our proposed systems `RPL' and `RPL+AL' with those of other state-of-the-art methods.
As mentioned in the introduction, our main goal is to propose the improved DML loss function instead of focusing on audio--text alignment. Therefore, we avoid comparing our work with works mainly dealing with audio--text alignment \cite{Shin-INTERSPEECH, Nishu-INTERSPEECH, Lee-INTERSPEECH}.

The first baseline method, called PATN \cite{Lim-arxiv}, is based on audio enrollment, using only the acoustic encoder, as explained in Section \ref{section:aux_losses}. 
It is noteworthy that our proposed methods achieve better performance than the audio enrollment approach, without the need for users to endure the cumbersome process of uttering specific keywords during enrollment.
The second and third baseline methods, the AsyP \cite{Jung-INTERSPEECH} and the AdaMS \cite{Jung-INTERSPEECH-AdaMS}, are explained in Section \ref{section:AsyP}. 
It is evident that RPL outperforms the baseline systems, and the addition of auxiliary losses leads to further improvements. `RPL+AL' attains a relative improvement of 22.96\% in terms of EER compared to PATN and 15.71\% in terms of AP compared to AsyP, respectively.

\section{Conclusion}
We have introduced the Relational Proxy Loss (RPL) for audio--text based keyword spotting (KWS). 
Unlike previous works using deep metric learning that only focused on comparing acoustic embedding (AE) and text embedding (TE) in a point-to-point manner, RPL exploits the structural relations within AEs and within TEs with respect to distance and angle.
Specifically, we introduced three variants of RPL, namely RPL-D, RPL-A, and RPL-P.
Our experiments demonstrated that combining point-to-point and structure-to-structure approaches led to better performance.
Further improvement was achieved by incorporating auxiliary losses.
On the noisy and reverberant WSJ test set, the proposed RPL and RPL+AL outperformed existing techniques, including both speech enrolled KWS and text enrolled KWS methods.
In the future, we will focus on audio--text alignment and try to integrate our RPL into an audio--text alignment framework that predominantly adopts a naive loss function.

\bibliographystyle{IEEEtran}
\bibliography{mybib}

\begin{thebibliography}{10}
\providecommand{\url}[1]{#1}
\csname url@samestyle\endcsname
\providecommand{\newblock}{\relax}
\providecommand{\bibinfo}[2]{#2}
\providecommand{\BIBentrySTDinterwordspacing}{\spaceskip=0pt\relax}
\providecommand{\BIBentryALTinterwordstretchfactor}{4}
\providecommand{\BIBentryALTinterwordspacing}{\spaceskip=\fontdimen2\font plus
\BIBentryALTinterwordstretchfactor\fontdimen3\font minus \fontdimen4\font\relax}
\providecommand{\BIBforeignlanguage}[2]{{%
\expandafter\ifx\csname l@#1\endcsname\relax
\typeout{** WARNING: IEEEtran.bst: No hyphenation pattern has been}%
\typeout{** loaded for the language `#1'. Using the pattern for}%
\typeout{** the default language instead.}%
\else
\language=\csname l@#1\endcsname
\fi
#2}}
\providecommand{\BIBdecl}{\relax}
\BIBdecl

\bibitem{Chen14-ICASSP}
G.~Chen, C.~Parada, and G.~Heigold, ``Small-footprint keyword spotting using deep neural networks,'' in \emph{Proc. IEEE International Conference on Acoustics, Speech and Signal Processing (ICASSP)}, 2014, pp. 4087--4091.

\bibitem{Sainath-INTERSPEECH}
T.~N. Sainath and C.~Parada, ``Convolutional neural networks for small-footprint keyword spotting,'' in \emph{Proc. Interspeech}, 2015, pp. 1478--1482.

\bibitem{TANG17-ICASSP}
R.~Tang and J.~J. Lin, ``Deep residual learning for small-footprint keyword spotting,'' in \emph{Proc. IEEE International Conference on Acoustics, Speech and Signal Processing (ICASSP)}, 2017, pp. 5484--5488.

\bibitem{Chen-ICASSP}
G.~Chen, C.~Parada, and T.~N. Sainath, ``Query-by-example keyword spotting using long short-term memory networks,'' in \emph{Proc. IEEE International Conference on Acoustics, Speech and Signal Processing (ICASSP)}, 2015, pp. 5236--5240.

\bibitem{Huang-ICASSP}
J.~Huang, W.~Gharbieh, H.~S. Shim, and E.~Kim, ``Query-by-example keyword spotting system using multi-head attention and soft-triple loss,'' in \emph{Proc. IEEE International Conference on Acoustics, Speech and Signal Processing (ICASSP)}, 2021, pp. 6858--6862.

\bibitem{Kurmi-INTERSPEECH}
K.~R, V.~K. Kurmi, V.~Namboodiri, and C.~V. Jawahar, ``Generalized keyword spotting using {ASR} embeddings,'' in \emph{Proc. Interspeech}, 2022, pp. 126--130.

\bibitem{Lim-arxiv}
H.~Lim, Y.~Kim, Y.~Jung, M.~Jung, and H.~Kim, ``Learning acoustic word embeddings with phonetically associated triplet network,'' \emph{arXiv:1811.02736}, 2018.

\bibitem{He-ICLR}
W.~He, W.~Wang, and K.~Livescu, ``Multi-view recurrent neural acoustic word embeddings,'' in \emph{Proc. International Conference on Learning Representations (ICLR)}, 2017.

\bibitem{Jung-INTERSPEECH}
M.~Jung and H.~Kim, ``Asymmetric proxy loss for multi-view acoustic word embeddings,'' in \emph{Proc. Interspeech}, 2022, pp. 5170--5174.

\bibitem{Jung-INTERSPEECH-AdaMS}
M.~Jung and H.~Kim, ``{AdaMS}: Deep metric learning with adaptive margin and adaptive scale for acoustic word discrimination,'' in \emph{Proc. Interspeech}, 2023, pp. 3924--3928.

\bibitem{Shin-INTERSPEECH}
H.-K. Shin, H.~Han, D.~Kim, S.-W. Chung, and H.-G. Kang, ``Learning audio-text agreement for open-vocabulary keyword spotting,'' in \emph{Proc. Interspeech}, 2022, pp. 1871--1875.

\bibitem{Nishu-INTERSPEECH}
K.~Nishu, M.~Cho, and D.~Naik, ``Matching latent encoding for audio-text based keyword spotting,'' in \emph{Proc. Interspeech}, 2023, pp. 1613--1617.

\bibitem{Lee-INTERSPEECH}
Y.-H. Lee and N.~Cho, ``{PhonMatchNet}: Phoneme-guided zero-shot keyword spotting for user-defined keywords,'' in \emph{Proc. Interspeech}, 2023, pp. 3964--3968.

\bibitem{Wang-CVPR}
X.~Wang, X.~Han, W.~Huang, D.~Dong, and M.~R. Scott, ``Multi-similarity loss with general pair weighting for deep metric learning,'' in \emph{Proc. the IEEE/CVF Conference on Computer Vision and Pattern Recognition (CVPR)}, 2019, pp. 5022--5030.

\bibitem{Park-CVPR}
W.~Park, D.~Kim, Y.~Lu, and M.~Cho, ``Relational knowledge distillation,'' in \emph{Proc. the IEEE/CVF Conference on Computer Vision and Pattern Recognition (CVPR)}, 2019, pp. 3967--3976.

\bibitem{Yi-ICPR}
D.~Yi, Z.~Lei, S.~Liao, and S.~Z. Li, ``Deep metric learning for person re-identification,'' in \emph{Proc. International Conference on Pattern Recognition}, 2014, pp. 34--39.

\bibitem{Huber-AMS}
P.~J. Huber, ``Robust estimation of a location parameter,'' \emph{The annals of mathematical statistics}, vol.~35, pp. 73--101, 1964.

\bibitem{Snell-NIPS}
J.~Snell, K.~Swersky, and R.~Zemel, ``Prototypical networks for few-shot learning,'' in \emph{Proc. Advances in Neural Information Processing Systems (NIPS)}, 2017.

\bibitem{Speechocean-DB}
\BIBentryALTinterwordspacing
DataOceanAI, ``King-{ASR}-066,'' 2015. [Online]. Available: \url{https://en.speechocean.com/datacenter/details/1446.html}
\BIBentrySTDinterwordspacing

\bibitem{McAuliffe-INTERSPEECH}
M.~McAuliffe, M.~Socolof, S.~Mihuc, M.~Wagner, and M.~Sonderegger, ``Montreal forced aligner: trainable text-speech alignment using kaldi,'' in \emph{Proc. Interspeech}, 2017, pp. 498--502.

\bibitem{Ko-ICASSP}
T.~Ko, V.~Peddinti, D.~Povey, M.~L. Seltzer, and S.~Khudanpur, ``A study on data augmentation of reverberant speech for robust speech recognition,'' in \emph{Proc. IEEE International Conference on Acoustics, Speech and Signal Processing (ICASSP)}, 2017, pp. 5220--5224.

\bibitem{Paul-WSN}
D.~B. Paul and J.~M. Baker, ``The design for the wall street journal-based {CSR} corpus,'' in \emph{Proc. the Workshop on Speech and Natural Language}, 1992, pp. 357--362.

\bibitem{Povey-ASRU}
D.~Povey, A.~Ghoshal, G.~Boulianne, L.~Burget, O.~Glembek, N.~Goel, M.~Hannemann, P.~Motlicek, Y.~Qian, P.~Schwarz, J.~Silovsky, G.~Stemmer, and K.~Vesely, ``The kaldi speech recognition toolkit,'' in \emph{Proc. IEEE Workshop on Automatic Speech Recognition and Understanding (ASRU)}, 2011.

\bibitem{Snyder-arxiv}
D.~Snyder, G.~Chen, and D.~Povey, ``Musan: A music, speech, and noise corpus,'' \emph{arXiv:1510.08484v1}, 2015.

\bibitem{Hu-INTERSPEECH}
Y.~Hu, S.~Settle, and K.~Livescu, ``Multilingual jointly trained acoustic and written word embeddings,'' in \emph{Proc. Interspeech}, 2020, pp. 1052--1056.

\bibitem{Jung-ASRU}
M.~Jung, H.~Lim, J.~Goo, Y.~Jung, and H.~Kim, ``Additional shared decoder on siamese multi-view encoders for learning acoustic word embeddings,'' in \emph{Proc. IEEE Workshop on Automatic Speech Recognition and Understanding (ASRU)}, 2019.

\bibitem{Desplanques-arxiv}
B.~Desplanques, J.~Thienpondt, and K.~Demuynck, ``{ECAPA-TDNN}: Emphasized channel attention, propagation and aggregation in {TDNN} based speaker verification,'' \emph{arXiv:2005.07143}, 2020.

\bibitem{Sennrich-ACL}
R.~Sennrich, B.~Haddow, and A.~Birch, ``Neural machine translation of rare words with subword units,'' in \emph{Proc. the 54th Annual Meeting of the Association for Computational Linguistics (ACL)}, 2016, pp. 1715--1725.

\bibitem{MoiHuggingFace-Tokenizers}
\BIBentryALTinterwordspacing
A.~Moi and N.~Patry, ``{HuggingFace's Tokenizers},'' 2023. [Online]. Available: \url{https://github.com/huggingface/tokenizers}
\BIBentrySTDinterwordspacing

\end{thebibliography}

\end{document}